\documentstyle[seceq,epsf,wrapft]{ptptex}
\newcommand{\bracket}[3]{{\left\langle {#1} \, \right|}{ {#2} \left| {#3} \,\right\rangle}}
\newcommand{\ket}[1]{\left| {#1}\,\right\rangle} 

\newcommand{\rbracket}[3]{{\left\langle {#1} \, \right\|}{ {#2} \left\| {#3} \,\right\rangle}}
\newcommand{\CG}[2]{ \left\langle {#1} \, | {#2} \,\right\rangle   } 
\nofigureboxrule
\notypesetlogo  
\title{
A New Description of Nuclear Rotational Motion\\
	in terms of Intrinsic Pair Mode }

\author{
Daisuke {\sc Hayashi}, Makoto {\sc Ueno}\footnote{Present address, FUJISOFT ABC Inc. 6-26 Nishiki 1-chome, Naka-ku, Nagoya-shi, Aichi 460-0003, Japan.},
 and Yoshinao {\sc Miyanishi}
}

\inst{
Graduate School of Science, Nagoya University, Nagoya 464-8602}

\recdate{
}

\abst{
A new method describing nuclear rotational motion microscopically is proposed. We extract the rotational Hamiltonian by introducing the intrinsic pair modes which commute with the rotational mode. Thereby the rotational mode is not treated as zero energy mode in contrast with the conventional RPA formalism so that we circumvent the difficulty related with infrared divergence. The wave function is constructed by angular momentum projection on each intrinsic state. Without numerical integration for projection we calculate the matrix elements analytically under a certain approximation. The numerical calculations are carried out to illustrate the applicability of our method and they show that our method works well.
}

\begin{document}
\maketitle

\section{Introduction}
The study of rotating nuclei is one of the main topics of the nuclear physics. In order to describe nuclear rotational motion microscopically, {\it the mean field approximation} is often used, such as {\it the cranked-Hartree-Fock-Bogoliubov} (CHFB) {\it theory} \cite{RS}. Although this approach succeeded in explaining the various phenomena of rotating nuclei, it has a serious defect; the method inevitably accompanies {\it the spontaneous symmetry breaking} and thus the angular momentum is conserved only on the average. The resultant wave functions are no more the eigenstates of angular momentum. 

Many works toward the restoration of broken symmetry have been made. One of the useful approaches is {\it the projection method} \cite{HI79}$^{-}$\cite{HI84}, that is {\it the angular momentum projection} (AMP) in this case. AMP is a mathematically rigorous method. However it needs much of a numerical effort. Furthermore the conventional treatment of AMP loses the simple picture of intrinsic excitation. Therefore the projection method has been applied in approximate version \cite{K68}\cite{NZ64}. Owing to the recent development of computer technology, the calculations with full projection, that is {\it the projected shell model} (PSM) {\it calculations} (for a review, see Ref.~\citen{HY95}, has been carried out although they are still limited. The application to axially deformed nuclei are carried out \cite{SE94}\tocite{LS01}. Furthermore the calculation including triaxiality have been made \cite{SH99}\tocite{SSP01} in transitional region. In these works {\it the Nilsson-plus-BCS state} is adopted as the intrinsic wave function, that means the self-consistent treatment of the coupling between the rotational mode and intrinsic mode is not fully considered. The application to the CHFB state, in which the axial symmetry is no more kept, is also made \cite{ETY98}\tocite{ETY992}, but the effect of the intrinsic excitation is not took into account. 

{\it The random-phase approximation} (RPA) is another useful approach. This method separates the rotational mode as zero-energy mode from the intrinsic mode and restores the broken symmetry within its approximation. However this method also has a shortcoming \cite{MW69}; it is valid for only the small amplitude oscillation, while the zero energy mode ranges over the large amplitude because of no restoring force. This discrepancy, for example, appears as the divergence of the norm of RPA ground state. Although the attempts to overcome this difficulty have been made, the description of finite rotation starting from the small amplitude approximation has its own difficulty \cite{MW69}\tocite{M76-2}.

In this work we propose a new method to get rid of the difficulty of RPA and to give an approximate treatment for AMP at the same time. The essential point is that we do not treat the angular momentum as an eigenmode although the intrinsic mode is decoupled with it. The wave function having the definite angular momentum is formally given by the angular momentum projection. However we give an analytic expression for the matrix elements between these states without integration. For simplicity, in the present work, the nucleus is assumed to be axially deformed.

In \S2, the outline of our method is shown together with a brief review of RPA. \S3 is devoted to the formulation of our method in which the rotational Hamiltonian is extracted by introducing the intrinsic operator. The wave functions are obtained in the subspace spanned by the multi-intrinsic pair base. In \S4, we derive the analytic expression for the moment of inertia and $B(E2)$ transition probability under a certain approximation. The numerical calculation is made to check the validity of our method and the results are compared with those of RPA and AMP in \S5. Finally the summary is given in \S6.

\section{Basic Idea}
We begin this section with a brief review of RPA. In axially symmetric system, $\hat J_z$ has the good quantum number. The RPA vibrational phonon operators ${\mathcal O}^\dagger_n,{\mathcal O}_n$ are derived from the equation
\begin{eqnarray}
	\left[ \hat H, {\mathcal O}^\dagger_n \right]_{RPA} &=& E_n {\mathcal O}^\dagger_n.
\label{eq:RPAeq}
\end{eqnarray}
The subscription $RPA$ denotes that the only leading terms are considered. In the solution of this equation, however, the zero energy mode $\hat J_\pm$ appears 
\begin{eqnarray}
	&&\left[ \hat H, \hat J_\pm \right]_{RPA} = 0,
\label{eq:Zmode}
\end{eqnarray}
where $\hat J_\pm = \hat J_x \pm i \hat J_y$. Eq.$(\ref{eq:Zmode})$ means that the symmetry broken by the mean field approximation is restored. $\hat H_{RPA}$ is completely divided into the rotational part and vibrational part within its approximation.
\begin{eqnarray}
	\hat H_{RPA} = E_{RPA} + \frac{\hat J_- \hat J_+}{2{\mathcal J}_0} + \sum_n E_n {\mathcal O}^\dagger_n {\mathcal O}_n ,
\label{eq:RPAH}
\end{eqnarray}
where $E_{RPA} = \bracket{0_{RPA}}{\hat H}{0_{RPA}}$ and $\ket{0_{RPA}}$ is RPA vacuum defined by
\begin{eqnarray}
	{\mathcal O}_n \ket{0_{RPA}} =  0, \\
	\hat J_\pm \ket{0_{RPA}} = 0.
\end{eqnarray}
However this definition leads to the well-known difficulty; the norm $\left< 0_{RPA} | 0_{RPA} \right>$ diverges infinitely. It stems from the fact the rotational mode is treated as an eigenmode.

Our approach is proposed in order to avoid this difficulty. We also separate the rotational mode and intrinsic mode but we never treat the rotational mode as an eigenmode. For the first step, the intrinsic pair operator $X^\dagger_n$ is defined as the commutable mode with $ \hat J_\pm$ in lowest order,
\begin{eqnarray}
	\left[ \hat J_\pm, X^\dagger_n \right] =  0.
\label{eq:intr}
\end{eqnarray}
The second step is to represent the Hamiltonian in terms of $\hat J_\pm, X^\dagger_n, X_n $ and to decompose it to three parts,
\begin{eqnarray}
	\hat H = \hat H_{rot} + \hat H_{intr} + \hat H_{coupl} + (\mbox{const.}),
\label{eq:h2}
\end{eqnarray}
where
\begin{eqnarray}
	\hat H_{rot}&=& \hat H_{rot}(\hat J_- \hat J_+),\\
	\hat H_{intr}&=&\hat H_{intr}( X^\dagger_n, X_n ),\\
	\hat H_{coupl}&=&\hat H_{coupl}(\hat J_\pm, X^\dagger_n,X_n).
\end{eqnarray}
For the last step, $\hat H$ is diagonalized in the subspace spanned by {\it the projected multi-intrinsic pair base} 
\begin{eqnarray}	
	\{ \ket{IM} \}= \{ \hat P^I_{M0}\ket{0},\hat P^I_{MK} X_n^\dagger \ket{0},\hat P^I_{MK}  X_n^\dagger  X_{n'}^\dagger \ket{0},...\}, 
\label{eq:MIPB}
\end{eqnarray}
where $\hat P^I_{MK}$ is the angular momentum projection operator.

In the usual projection method like PSM, $\hat H$ is diagonalized in {\it the projected Tamm-Dancoff base} 
\begin{eqnarray}	
	\{ \ket{IM} \}= \{\hat P^I_{M0}\ket{0},\hat P^I_{MK}  \left( a^\dagger a^\dagger\right)_\pi \ket{0},\hat P^I_{MK} \left( a^\dagger a^\dagger\right)_\nu \ket{0},...\}. 
\end{eqnarray}
It should be noted that $(a^\dagger a^\dagger)_\pi, ( a^\dagger a^\dagger)_\nu$ include both the rotational and intrinsic modes, while the rotational mode is already eliminated in $X_n^\dagger$ in lowest order. This leads to an advantage of our approach that the analytic form of matrix is approximately obtained without integral calculation. (See \S3.)
	
\section{Formulation}
	\subsection{Intrinsic Pair Operator}
In axially symmetric system, any operator is classified by $K$ quantum number. In this case, $\hat J_\pm$ corresponds to the rotational mode. It is written in the quasi-particle representation as
\begin{eqnarray}
		\hat J_\pm &=& \sum_{\alpha < \beta }
 \left\{ 
	 J^{20}_{\pm \: \alpha \beta} a^{\dagger}_{\alpha} a^{\dagger}_{\beta}
	+J^{02}_{\pm \: \alpha \beta} a_{\beta}a_{\alpha} 
\right\}
+\sum_{\alpha \beta } 
	J^{11}_{\pm \: \alpha \beta} a^{\dagger}_{\alpha} a_{\beta},
\end{eqnarray}
where we follow the notation of Ref.~\citen{RS}. The collective pair operator $X^\dagger_\pm$ is defined as
\begin{eqnarray}
	X_\pm^\dagger &\equiv& \sum_{\alpha < \beta} \psi^{({\pm})}_{\alpha \beta} a^{\dagger}_{\alpha} a^{\dagger}_{\beta},\\
	\hat J_\pm &=& a_0\left( X_\pm^\dagger + X_\mp \right)+ \sum_{\alpha \beta} J^{11}_{\pm \: \alpha \beta} a^{\dagger}_{\alpha} a_{\beta},
\end{eqnarray}
where $a_0=\bracket{0}{\hat J_- \hat J_+}{0}^{1/2}$. Since $a_0$ is large enough in deformed nuclei
\begin{eqnarray}
	\hat J_\pm \simeq a_0\left( X_\pm^\dagger + X_\mp \right)
\label{eq:Bapprox}
\end{eqnarray}
The intrinsic pair operator is defined so as it may satisfy $(\ref{eq:intr})$ up to the leading order, that means 
\begin{eqnarray}
	&X_n^{\dagger} \equiv \sum_{\alpha < \beta} \psi^{(n)}_{\alpha \beta}
 a^{\dagger}_{\alpha} a^{\dagger}_{\beta}\\
	& \sum_{\alpha < \beta }{\psi}^{(\pm)*}_{\alpha \beta}{\psi}^{(n)}_{\alpha \beta}= 0.
\end{eqnarray}

From now on, we use the following notation rule; the subscription '$\pm$' or '$c$' denotes the collective mode and '$n$' denotes the intrinsic one and '$\mu,\nu,...$' denotes both. If necessary, $K$ quantum number subscription is added, as {\it e.g. $X^{\dagger}_{n K}$}. The amplitudes ${\psi}^{(\mu)}_{\alpha \beta}$ satisfy the orthogonality and completeness
\begin{eqnarray}
	\sum_{\alpha < \beta }{\psi}^{(\mu)*}_{\alpha \beta}{\psi}^{(\nu)}_{\alpha \beta}&=& \delta_{\mu \nu},\\
\sum_{\mu }{\psi}^{(\mu)*}_{\alpha \beta}{\psi}^{(\nu)}_{\alpha' \beta'}&=& \delta_{\alpha \alpha'} \delta_{\beta \beta'} \qquad ({\alpha < \beta }).
\end{eqnarray}
They can be simultaneously determined from the eigenvalue equation for the Stability matrix 
\begin{eqnarray}
	\sum_{\alpha < \beta }\sum_{\alpha' < \beta' }\psi^{(\mu)*}_{\alpha \beta}( \mathcal{A}_{\alpha \beta ;\alpha' \beta'} + \mathcal{B}_{\alpha \beta ;\overline{\alpha' \beta'} } ) 
\psi^{(\nu)}_{\alpha' \beta'} = \omega_{\mu} \delta_{\mu \nu},
\label{eq:diag}
\end{eqnarray}
where
\begin{eqnarray}
	\mathcal{A}_{\alpha \beta ;\alpha' \beta'}&=& \bracket{0}{[a_{\beta}a_{\alpha},[\hat H,a^{\dagger}_{\alpha'} a^{\dagger}_{\beta'}]]}{0},\\
	\mathcal{B}_{\alpha \beta ;\alpha' \beta'}&=& -\bracket{0}{[a_{\beta}a_{\alpha},[\hat H,a_{\alpha'} a_{\beta'}]]}{0}, 
\end{eqnarray}
and $\overline{\alpha' \beta'}$ denotes the time-reversal index of $\alpha' \beta'$.	

	\subsection{Extraction of the rotational Hamiltonian}
The Hamiltonian with two-body interaction is written in terms of pair operators up to leading order (Appendix A),
\begin{eqnarray}
	\hat H 	&\simeq& \langle{H} \rangle 
		+\sum_n \omega_n  X_n^{\dagger} X_n +\sum_{\mu\nu}\left[ V_{v (\mu \bar{\nu})} \left(X_\mu^\dagger X_{\bar \nu}^\dagger+ X_{\bar \nu}X_\mu +2X_\mu^\dagger X_\nu\right) \right] ,\quad \label{eq:ham}
\end{eqnarray}
where $V_{v (\mu \nu)}$ is a newly defined matrix
\begin{eqnarray}
	V_{v (\mu \nu)}&=&\sum_{\alpha  \beta }\sum_{\alpha'  \beta' }
	         \psi^{(\mu)*}_{\alpha \beta} H^{40}_{\alpha \beta ;\alpha' \beta'}\psi^{(\nu)*}_{\alpha' \beta'}.
\end{eqnarray}
By classifying $(\ref{eq:ham})$ with respect to each mode, $\hat H$ can be decomposed into the rotational part $\hat H_{rot}$ which can be written only by the collective pairs, the intrinsic part $\hat H_{intr}$ only by the intrinsic pairs, and their coupling part $\hat H_{coupl}$
\begin{eqnarray}
	\hat H_{rot}&=&\frac{ 2V_{v(+-)}}{a_0^2}\hat J_- \hat J_+\label{eq:rot},\\
	\hat H_{intr}&=&\sum_n \omega_n X_n^{\dagger} X_n +\sum_{n,n'} V_{v (n \bar{n'})} 
\left( X_n^\dagger X_{\bar{n'}}^\dagger+ X_{\bar{n'}}X_n +2X_n^{\dagger}X_{n'}\right), \\
	\hat H_{coupl}&=&  2\sum_{n,c} V_{v (n \bar c)}\left( X_n^\dagger X_{\bar c}^\dagger+ X_{\bar c}X_n +X_n^{\dagger}X_{c}+X_{\bar c}^{\dagger}X_{\bar n}\right).
\label{eq:h3}
\end{eqnarray}
In Eq.$(\ref{eq:rot})$ we used $(\ref{eq:Bapprox})$.

	\subsection{Wave function}
We span the intrinsic space
\begin{eqnarray}	
	\ket{k_l} &=& X^\dagger_{n_1}...X^\dagger_{n_l}\ket{0},
\label{eq:space}
\end{eqnarray}
and make the wave function in the following base 
\begin{eqnarray}	
	\{ \ket{IM} \}&=& \{ \hat P^I_{MK}\ket{k_l}\}
\end{eqnarray}
where
\begin{eqnarray}	
	\hat P^I_{MK}&=&\frac{2I+1}{8\pi^2}\int^{2\pi}_0 d\alpha \int^\pi_0 d\beta sin\beta  \int^{2\pi}_0 d\gamma D^{I^*}_{MK}(\alpha, \beta, \gamma) e^{-i\alpha \hat J_z} e^{-i\beta \hat J_y} e^{-i\gamma \hat J_z}.\nonumber\\
\end{eqnarray}
Because $\hat J_{\pm}$ change into c-numbers when they are operated to these states, all we have to do is calculate the norm matrix of this base. 

Here we note that $K \ge 0$ intrinsic pairs have to be prepared to compose the orthogonal base because
\begin{eqnarray}
	\bracket{0}{X_{n K} P^I_{KK}X^\dagger _{nK}}{0}
	&=& \bracket{0}{X_{n -K} P^I_{-KK}X^\dagger _{nK}}{0}\nonumber\\
	&=& \bracket{0}{X_{n -K} P^I_{-K-K}X^\dagger _{n-K}}{0}.
\end{eqnarray}
This property stems from the time-reversal symmetry of the HFB wave function in the axially deformed system. Moreover, through this study, we neglect the non-diagonal elements of the norm matrix
\begin{eqnarray}
	\bracket{0}{\hat P^I_{0K}X^\dagger_{n_K}}{0}, \quad \bracket{0}{\hat P^I_{MK}X^\dagger_{n} X^\dagger_{n'}}{0}, \quad \bracket{0}{X_n  \hat P^I_{MK}X^\dagger_{n'} X^\dagger_{n''}}{0},...
\label{eq:Ndiag}
\end{eqnarray}
compared with $\bracket{0}{\hat P^I_{00}}{0}$. This is because $\hat P^I_{MK}$ includes only the rotational mode which is orthogonal to the intrinsic mode in lowest order. Under this approximation, we can calculate the matrix without integration.

\section{Physical Quantities}
	\subsection{Moment of Inertia}
		\subsubsection{Peierls-Yoccoz' Moment of Inertia}
{\it The Peierls-Yoccoz' moment of inertia}\cite{Y57} is defined 
\begin{eqnarray}
	E_I &=& \bracket{IM}{\hat H}{IM}\nonumber\\
	    &=& E_0 +  \frac{ I(I+1) }{ 2{\mathcal{J}}_{PY}}, 
\label{eq:PY}
\end{eqnarray}
where $\ket{I M}$ is determined as following
\begin{eqnarray}
	\ket{IM}&=& {\mathcal N}_0^I\hat P^I_{M0}\ket{0}
\end{eqnarray}
where the normalization constant ${\mathcal N}^I_0=\bracket{0}{P^I_{00}}{0}^{-1/2}$. Comparing $(\ref{eq:PY})$ with $(\ref{eq:rot})-(\ref{eq:h3})$ under the approximation mentioned above, the Peierls-Yoccoz' moment of inertia can be derived analytically
\begin{eqnarray}
	{\mathcal J}_{PY} = \frac {a_0^2}{ 4 Vv_{(+-)} }.
\end{eqnarray}

		\subsubsection{Thouless-Valatin Moment of Inertia}
As a rule {\it the Peierls-Yoccoz' moment of inertia} tends to underestimate the experimental value\cite{VDS83}, and thus we extend the physical space by adding $K=1$ one phonon states perturbatively
\begin{eqnarray}	
	\ket{IM} &\simeq&  {\mathcal N}^I_0 \hat P^I_{M 0}\ket{0} +\eta \sum_{n_{K=+1}}  \; {\mathcal N}^I_{n_{+1}} \hat P^I_{M1} X^\dagger_{n_{+1}}\ket{0},
\end{eqnarray}
where
\begin{eqnarray}	
	{\mathcal N}^I_{n_K} &=& \bracket{0}{X_{n_K} \hat P^I_{KK}X_{n_K}^\dagger }{0}^{-1/2}.\nonumber
\end{eqnarray}
The perturbative energy shift in second order is given by 
\begin{eqnarray}
	\Delta E &=&  -\sum_{n_{K=+1}} \frac{|{\mathcal N}^I_0 {\mathcal N}^I_{n_{+1}} \bracket{0}{\hat P^I_{0M} \hat H \hat P^I_{M1}X_{n_{+1}}^\dagger }{0} |^2}{\omega_n} .\label{eq:Esf}
\end{eqnarray}
Here we must mention that the definitions of $\omega_n$ and $ X_{n_K}^\dagger$ are slightly different from those of $(\ref{eq:diag})$ because the formers are defined as the Tamm-Dancoff eigenvalues and eigenvectors in the intrinsic space, i.e. they are derived by  diagonalization of ${\mathcal A}$ after removing the zero energy modes. 

In addition to $(\ref{eq:Ndiag})$, we also neglect the terms 
\begin{eqnarray}
	\bracket{0}{X_{n_{+1}}\hat P^I_{11}X^\dagger_{n'_{+1}}}{0}. \qquad (n \ne n')
\end{eqnarray}
compared with $\bracket{0}{\hat P^I_{00}}{0}$. $\Delta E$ can be derived analytically
\begin{eqnarray}
	\Delta E   &\simeq& -8I(I+1)  a_0^{-2} \sum_{n_{K=+1}} \omega^{-1}_n V_v^2(n_{+1}\;-).
\end{eqnarray}
Here we used the approximate relation (see Appendix B),
\begin{eqnarray}
	\frac {\bracket{0}{X_{n_{+1}} \hat P^I_{11}X_{n_{+1}}^\dagger }{0} }
	{ \bracket{0}{\hat P^I_{0 0}}{0} } 
		\simeq \frac {V_v(+ \; n_{-1} )}{2 V_v(n_{+1}\; -)}= \frac{1}{2}.
\label{eq:approx1}
\end{eqnarray}
which can be naturally derived from our approximation. Thereby the moment of inertia is newly defined  
\begin{eqnarray}
	\frac{ I(I+1) }{ 2{\mathcal J}_{PY}} + \Delta E = \frac{I(I+1)}{2 {\mathcal J}_{eff}}.	
\end{eqnarray}
where ${\mathcal J}_{eff}$ corresponds to {\it the Thouless-Valatin moment of inertia} ${\mathcal J}_{TV}$, which coincides with ${\mathcal J}_0$ in $(\ref{eq:RPAH})$. It is calculated from
\begin{eqnarray}
	{\mathcal J}_0 = 2 \sum_{\alpha \beta;\alpha' \beta'} 
		\psi^{(c)*}_{\alpha \beta}( {\mathcal A-B} )^{-1}_{\alpha \beta;\alpha' \beta'} \psi^{(c)}_{\alpha' \beta'} .
\end{eqnarray}

	\subsection{B(E2) Transition Probability}
The $B(E2)$ transition probability in the ground band also derived in the same framework. In the space $\ket{IM}={\mathcal N}_0 \hat P^I_{M0}\ket{0}$, it can be written 
\begin{eqnarray}
	B(E2;I+2 \rightarrow I)= \left[ \frac{ {\mathcal N}_0^{I+2} }{{\mathcal N}_0^I} \sum_{\lambda=-2}^2 \CG{I\:-\lambda;2\:\lambda}{I+2\:0} \; g_\lambda (I)\right]^2, 
\end{eqnarray}
where $g_\lambda (I)$ is given by
\begin{eqnarray}
	g_\lambda (I)\equiv (-1)^\lambda \frac{ \bracket{0}{ \hat P^I_{0\lambda}\hat Q_{2\lambda} }{0} }{ \bracket{0}{ \hat P^I_{00} }{0} }.
\end{eqnarray}
$g_\lambda (I)$ and ${\mathcal N}^{I+2}_0/{\mathcal N}^I_0$ can be given the analytic form in our approximation. (See Appendix C.)
\begin{equation}
	g_\lambda (I) \simeq  \left\{
\begin{array}{ll}
	\left \langle \hat Q_{20} \right \rangle & (\mbox{for} \quad\lambda=0)\\
	-\xi_\pm a_0^{-1} \sqrt{I(I+1)} & (\mbox{for} \quad\lambda=\pm 1)\quad .\\
	 0 & (\mbox{for} \quad\lambda=\pm 2)
\end{array}\right.
\label{eq:glam}
\end{equation}
Here $\xi_\mu$ is the overlap coefficient 
\begin{eqnarray}
	\xi_{\mu_K} &=& \bracket{0}{X_{\mu_K} \hat Q_{2 K}}{0}
\end{eqnarray}
and
\begin{eqnarray}
\left( \frac{ {\mathcal N}^{I+2}_0}{{\mathcal N}^I_0 } \right)^2 =
	 \frac{ \bracket{0}{\hat P^I_{00} }{0}}  { \bracket{0}{ \hat P^{I+2}_{00}}{0}} 
= \sqrt{ \frac{2I+1}{2I+5} }\quad	
	\frac{ \sum_\lambda \CG{I\!+\!2\;\lambda ;2 \; -\lambda}{I0}\: g_{\lambda (I+2)} } 
	{ \sum_\lambda \CG{I\;\lambda ;2\; -\lambda}{I\!+\!2\;0}\: g_{\lambda (I)} }.\nonumber\\
\label{eq:Nratio}
\end{eqnarray}
From $(\ref{eq:Nratio})$ and the normalization condition $\sum_I \bracket{0}{\hat P^I}{0}=1 $, the angular momentum distributions of the HFB wave function $\ket{0}$ can be analytically evaluated.

\section{Numerical Calculation}
We carry out the numerical calculation applying our method to Er isotopes and compare the results with AMP and RPA. As the purpose is to check the validity of our method, we adopt a rather simple effective interaction; {\it the Pairing-plus-Quadrupole interaction}. The detail correspondence with experimental data by using the more realistic interaction is left in the future study.

The adopted Hamiltonian is 
\begin{eqnarray}
	\hat H&=& \hat H_0 -\frac{1}{2} \chi \sum _{\lambda =-2}^2 \hat Q_{2\lambda}^\dagger \hat Q_{2\lambda} 
	-G \hat P^\dagger \hat P, 
\end{eqnarray}
where the chemical potentials are already absorbed in the spherical single particle energy part $\hat H_0$. The single particle basis employed is $N$ = 4,5 and 6 for neutrons (64 levels) and $N$ = 3,4 and 5 for protons (46 levels). The exchange terms are neglected. The force parameters are below; for the {\it Q-Q interaction}
\begin{eqnarray}
	\chi_{nn} &=& 3.768 \times 10^{-2} \quad (\mbox{MeV}),\\
	\chi_{pp} &=& 4.796 \times 10^{-2} \quad (\mbox{MeV}),\\
	\chi_{np} &=& \chi_{pn} = \sqrt{ \chi_{nn}\chi_{pp} }\quad ,
\end{eqnarray}
where $\hat Q$ is scaled to be dimensionless by the oscillator length $b_0 = \sqrt{\frac{\hbar}{m\omega_0}}$. For {\it the pairing interaction} we use
\begin{eqnarray}
	G = (22.36 \mp 14.59 \frac{N-Z}{A} ) /A \quad (\mbox{MeV}),
\end{eqnarray}
where - for neutron and + for proton. \\

In table \ref{table:1} and II, the calculated results for the moment of inertia ${\mathcal J}_{PY}, {\mathcal J}_{T.V.}$ are shown. They indicate that our method gives a good approximation for AMP and RPA. 

\noindent
\begin{minipage}[t]{\halftext}
\begin{wraptable}{c}{5.5cm}
	\caption{Comparison of $ {\mathcal J}_{P.Y.} (\hbar^2 \cdot  \mbox{MeV}^{-1})$ obtained from AMP with the one from our method}
\label{table:1}
\begin{center}
\begin{tabular}{ccccc} \hline \hline
$ $ & ${}^{162}$Er  & ${}^{164}$Er & ${}^{166}$Er & ${}^{168}$Er    \\
\hline
AMP  &	20.6	& 22.7  & 24.5 & 26.8    \\
ours &	22.4 	& 24.5  & 26.1 & 28.6    \\
\hline
\end{tabular}
\end{center}
\end{wraptable}
\end{minipage}
$\quad$
\begin{minipage}[t]{\halftext}
\begin{wraptable}{c}{5.5cm}
	\caption{Comparison of $ {\mathcal J}_{T.V.} (\hbar^2 \cdot  \mbox{MeV}^{-1})$ obtained from RPA with the one from our method}
\label{table:2}
\begin{center}
\begin{tabular}{ccccc} \hline \hline
$ $ & ${}^{162}$Er & ${}^{164}$Er  & ${}^{166}$Er & ${}^{168}$Er  \\
\hline
RPA	& 29.8 & 33.1  & 35.0 & 38.0 	\\
ours	& 31.3 & 34.9  & 36.8 & 39.8 	\\
\hline
\end{tabular}
\end{center}
\end{wraptable}
\end{minipage}
\vspace{0.5cm}

The small discrepancies among them are supposed to be the consequence of our approximation $(\ref{eq:Ndiag})$. In order to check the accuracy of the diagonal element $ \bracket{0}{X_{n_{+1}} \hat P^I_{11}X_{n_{+1}}^\dagger }{0} / \bracket{0}{\hat P^I_{0 0}}{0}$  $(\ref{eq:approx1})$, we calculated it by the exact AMP. The results are shown in Table \ref{table:3}. Here we showed the data for the lowest energy intrinsic mode. (We checked the results for the other mode are nearly same.)

\begin{figure}
	\epsfysize = 5.15cm
	\centerline{\epsfbox{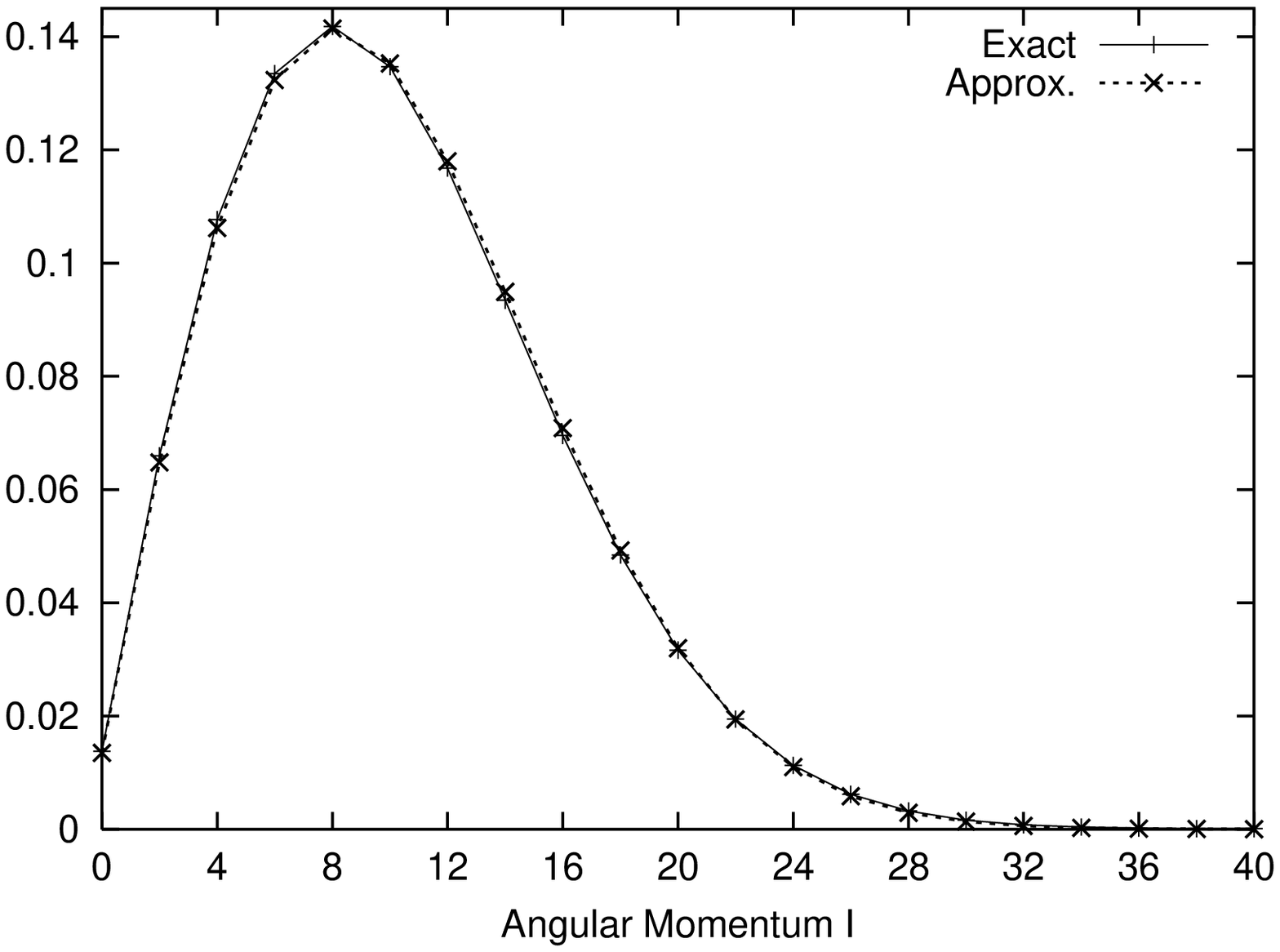}}
\caption{Angular momentum distribution  $\bracket{0}{\hat P^I}{0} $ for ${}^{166}$Er}
	\label{fig:1}
\end{figure}

The $B(E2)$ transition probability and the angular momentum distributions $\bracket{0}{\hat P^I}{0}$ for ${}^{166}$Er are shown in Table \ref{table:4} and Fig.\ref{fig:1}. In this calculation, we used the isoscalar effective charge $e_{eff}=0.3e$. Our method can well reproduce the result of AMP. To look into the origin of good agreements, we calculated $g_\lambda (I)$ by the exact AMP. The results are given in Table \ref{table:5}. We find that the main part of $B(E2)$ comes from $g_{\lambda=0}(I)$, which does not depend on the detail of the projection because it is mainly determined by $<Q_{20}>$. The same reason holds for the good agreement of $\bracket{0}{P^I_{00}}{0}$ with the exact one.

\begin{minipage}[t]{\halftext}
\begin{wraptable}{c}{5.0cm}
	\caption{ Comparison of Eq.$(\ref{eq:approx1})$ obtained from AMP with the one from our method } 
	\label{table:3}
\begin{center} 
\begin{tabular}[t]{ccc} \hline \hline
$I$ & AMP & approx.\\
\hline
2 & 0.49184  &		\\
4 & 0.49313  &		\\
6 & 0.49515  & 0.5	\\
8 & 0.49788  &		\\
10& 0.50129  &		\\
12& 0.50536  &		\\
\hline
\end{tabular}
\end{center}
\end{wraptable}

\end{minipage}
$\quad$
\begin{minipage}[t]{\halftext}
\begin{wraptable}{c}{5.5cm}
	\caption{B(E2;I+2 $\rightarrow$ I) ($e^2 \cdot b^2$) of ${}^{166}$Er obtained form exact AMP and our method.}
\label{table:4}
\begin{center}
\begin{tabular}[t]{ccc} \hline \hline
$I$ & \quad AMP \quad  & \quad ours \quad \\
\hline
 0  & 1.55  & 1.59  \\
 2  & 2.22  & 2.26  \\
 4  & 2.44  & 2.48  \\
 6  & 2.56  & 2.59  \\
 8  & 2.63  & 2.64  \\
10  & 2.68  & 2.67  \\
12  & 2.71  & 2.68  \\
\hline
\end{tabular}
\end{center}
\end{wraptable}

\end{minipage}
\begin{wraptable}{c}{\textwidth}
	\caption{ $g_\lambda (I)$ values obtained from exact AMP and our approximation}
\label{table:5}
\begin{center}
\begin{tabular}{c|cc|cc|cc} \hline  \hline 
 	&   \multicolumn{2}{|c}{ $g_{\lambda=0}(I)$  } & \multicolumn{2}{|c}{ $g_{\lambda=1}(I)$ }&  \multicolumn{2}{|c}{$g_{\lambda=2}(I)$}   \\
\hline
$I$ & exact &  approx. &   exact &  approx. &  exact &  approx. \\
\hline
 0 & 77.6 &	 	& 0	 & 0	& 0 	 &  	\\
 2 & 77.6 & 		& 1.62	 & 1.61 & 0.014 &	 \\
 4 & 77.7 & 		& 2.96	 & 2.94 & 0.053 &	 \\
 6 & 77.9 & 78.5	& 4.29	 & 4.26 & 0.113 &  0	 \\
 8 & 78.1 & 		& 5.61	 & 5.58 & 0.195 &  	 \\
10 & 78.3 & 		& 6.93	 & 6.90 & 0.297 &	 \\
12 & 78.6 & 		& 8.24	 & 8.21 & 0.419 &	 \\
\hline
\end{tabular}
\end{center}
\end{wraptable}
\section{Summary}
In this work, we proposed a new microscopic method for the description of nuclear rotational motion that avoids the difficulty of RPA. We separate the rotational Hamiltonian by defining the intrinsic mode that commutes with the rotational mode up to leading order. The wave function with definite angular momentum is formally determined through angular momentum projection, though practical integral calculation is replaced by the analytic estimation under a certain approximation. We calculated moment of inertia and $B(E2)$ transition probability in the ground band by using this method, restricting ourselves to axially deformed system. The obtained results shows good agreements with the ones from AMP and RPA.

This method has various advantages as following; (i) It keeps the clear physical picture of RPA treating the rotational and the intrinsic modes separately. (ii) all the necessary calculations are reduced to merely analytic ones and thus it is easier to include much of an intrinsic excitation. 

Finally we mention that this work is still preliminary. To compare with experimental data, more realistic interaction should be applied, such as {\it the pairing-plus-Quadrupole-plus-Quadrupole-pairing interaction, the Gogny interaction}. Moreover, more elaborate formulation that deals with $K$ quantum number mixing has to be made for the study of high spin state. These works are now in progress.

\section*{Acknowledgments}
This work is supported by many useful discussions of the Nagoya Nuclear Structure Seminar. We would like to thank to all the members.

\appendix
\section{Phonon Representation of Hamiltonian} 
The Hamiltonian with two-body interaction can be written in the quasi-particle space
\begin{eqnarray}
	\hat H =  \langle{\hat H} \rangle + \hat H^{11} + \frac{1}{2}(\hat H^{20} + \mbox{h.c.}) + \frac{1}{4}\hat H^{22} + (\hat H^{31}+ \mbox{h.c.}) +  ( \hat H^{40} + \mbox{h.c.}),\nonumber\\
\end{eqnarray}
where
\begin{eqnarray}
	\langle{\hat H} \rangle  &=& \bracket{0}{\hat H}{0},\\
	\hat H^{11}&=& \sum_{\alpha}E_{\alpha}\hat B_{\alpha \alpha},\\
	\hat H^{22}&=& \sum_{\alpha  \beta }\sum_{\alpha'  \beta' }H^{22}_{\alpha \beta ;\alpha' \beta'}a^{\dagger}_{\alpha}a^{\dagger}_{\beta}a_{\beta'} a_{\alpha'},\\
	\hat H^{40}&=& \sum_{\alpha  \beta }\sum_{\alpha'  \beta' }H^{40}_{\alpha \beta ;\alpha' \beta'}a^{\dagger}_{\alpha}a^{\dagger}_{\beta}a^{\dagger}_{\alpha'} a^{\dagger}_{\beta'},\\
	\hat H^{31}&=& \sum_{\alpha  \beta }\sum_{\alpha'  \beta' }H^{31}_{\alpha \beta ;\alpha' \beta'}a^{\dagger}_{\alpha}a^{\dagger}_{\beta}a^{\dagger}_{\alpha'} a_{\beta'},
\end{eqnarray}
and $\hat H^{20}=0$ in the HFB ground state. $E_\alpha$ is the quasi-particle energy and $\hat B_{\alpha \beta}$ is the scattering operator defined by
\begin{eqnarray}
	\hat B_{\alpha \beta} \equiv a^{\dagger}_{\alpha} a_{\beta},
\end{eqnarray}
Between $\hat B_{\alpha \beta}$ and the pair operators, there exist the following commutation relations 
\begin{eqnarray}
	 [ X_{\mu},X_{\nu}^{\dagger} ] &=& \delta_{\nu \mu}+\sum_{\alpha \beta}\Gamma^{(\nu \mu)}_{\alpha \beta}\hat B_{\alpha \beta},\\
	 {[} \hat B_{\alpha \beta},X_{\mu}^{\dagger}{]}  &=& - \sum_{\nu}\Gamma^{(\nu \mu)}_{\alpha \beta}X_{\nu}^{\dagger},
\label{eq:com}
\end{eqnarray}
where
\begin{eqnarray}
	\Gamma^{(\mu \nu)}_{\alpha \beta}&=& \sum_{\gamma}\psi^{(\mu)*}_{\alpha \gamma}\psi^{(\nu)}_{\gamma \beta}
\end{eqnarray}
is the first order of magnitude with respect to the small parameter $\epsilon\sim \frac{1}{\sqrt{\Omega}}$ ($\Omega$ is the number of states). $\hat B_{\alpha \beta}$ can be rewritten
\begin{eqnarray}
	\hat B_{\alpha \beta} &=&- \sum_{\mu \nu} \Gamma^{(\mu \nu)}_{\alpha \beta}X_{\mu}^{\dagger}X_{\nu} + O(\epsilon^2)
\end{eqnarray}
since this form satisfies $(\ref{eq:com})$ except for $O(\epsilon^2)$ deviation. Using the newly defined matrices
\begin{eqnarray}
	V_{x (\mu \nu)}&=&\sum_{\alpha  \beta }\sum_{\alpha'  \beta' }
	         \psi^{(\mu)*}_{\alpha \beta}H^{22}_{\alpha \beta ;\alpha' \beta'}\psi^{(\nu)}_{\alpha' \beta'},\\
	V_{v (\mu \nu)}&=&\sum_{\alpha  \beta }\sum_{\alpha'  \beta' }
	         \psi^{(\mu)*}_{\alpha \beta}H^{40}_{\alpha \beta ;\alpha' \beta'}\psi^{(\nu)*}_{\alpha' \beta'},\\
	V_{y (\mu;\alpha' \beta')}&=&\sum_{\alpha  \beta } \psi^{(\mu)*}_{\alpha \beta}H^{31}_{\alpha \beta ;\alpha' \beta'},
\end{eqnarray}
$\hat H$ is rewritten as
\begin{eqnarray}
	 \hat H &=& \langle{\hat H} \rangle \nonumber\\
	   &&+\sum_{\mu \nu} \left\{ \left(-\sum_{\alpha}E_{\alpha} \Gamma^{(\mu \nu)}_{\alpha \alpha}+\frac{1}{4}  V_{x (\mu \nu)} \right) X_{\mu}^{\dagger}X_{\nu}
		+ V_{v (\mu \bar{\nu})} \left( X_{\mu}^{\dagger}X_{\bar \nu}^{\dagger} + X_{\bar \nu} X_{\mu} \right) \right\} +O(\epsilon^2),\nonumber\\
\label{eq:hp}
\end{eqnarray}
As the Stability Matrix is (without exchange terms)
\begin{eqnarray} 
	{\mathcal A}_{\alpha \beta ; \alpha' \beta'} &=& (E_\alpha - E_\beta)(\delta_{\alpha \alpha'} \delta_{\beta \beta'}-\delta_{\alpha \beta'} \delta_{\beta \alpha'}) + H^{22}_{\alpha \beta ; \alpha' \beta'},\\
	{\mathcal B}_{\alpha \beta ; \alpha' \beta'} &=& 8 H^{40}_{\alpha \beta ; \alpha' \beta'} ,
\end{eqnarray}
$(\ref{eq:diag})$ can be rewritten
\begin{eqnarray}
	-\sum_{\alpha}E_{\alpha}\Gamma^{(\mu \nu)}_{\alpha \alpha}
	+ \frac{1}{4}V_{x (\mu \nu)} - 2V_{v (\mu \bar{\nu})} = \omega_{\mu} \delta_{\mu \nu}.
\label{eq:Sp}
\end{eqnarray}
From $(\ref{eq:hp})(\ref{eq:Sp})$, the phonon representation of $\hat H$ $(\ref{eq:ham})$ is derived.

\section{Evaluation of the $\Delta E$}
The perturbative energy shift in the second order is
\begin{eqnarray}
	\Delta E &=&  -\sum_{n_{K=1}} \frac{|{\mathcal N}^I_0 {\mathcal N}^I_{n_{+1}} \bracket{0}{\hat P^I_{0M} \hat H \hat P^I_{M1}X_{n_{+1}}^\dagger }{0} |^2}{\omega_n} .\label{eq:Esf}
\end{eqnarray}
From $\left[\hat H,\hat P^I_{MK} \right]=0$, 
\begin{eqnarray}
	\bracket{0}{\hat P^I_{0 M} \hat H \hat P^I_{M1}X_{n_{+1}}^\dagger }{0}
		&=& \bracket{0}{\hat H \hat P^I_{01}X_{n_{+1}}^\dagger }{0}\label{eq:ph1}\\
		&=& \bracket{0}{\hat P^I_{0K} \hat H X_{n_{+1}}^\dagger }{0}. \label{eq:ph2}
\end{eqnarray}
Calculating $(\ref{eq:ph1}),(\ref{eq:ph2})$ separately under our approximation,
\begin{eqnarray}
		(\ref{eq:ph1})&\simeq&\bracket{0}{\hat H_{coupl.} \hat P^I_{01}X_{n_{+1}}^\dagger }{0} \nonumber\\
		&\simeq&2 V_v(n_{+1}\; -)\bracket{0}{X_{n_{+1}}X_- \hat P^I_{01}X_{n_{+1}}^\dagger }{0} + 2 V_v(n_{-1}\; +)\bracket{0}{X_{n_{-1}}X_+ \hat P^I_{01}X_{n_{+1}}^\dagger }{0}
\nonumber\\
		&=& \sqrt{I(I+1)} \: 4 a^{-1}_0 V_v(n_{+1}\; -)\bracket{0}{X_{n_{+1}} \hat P^I_{11}X_{n_{+1}}^\dagger }{0},\label{eq:ph3}\\
\smallskip\nonumber\\
		(\ref{eq:ph2})&\simeq&\bracket{0}{\hat P^I_{01}\hat H_{coupl.}  X_{n_{+1}}^\dagger}{0}\nonumber\\
		&\simeq&\sqrt{I(I+1)} \: 2a^{-1}_0 V_v(+ \; n_{-1} )\bracket{0}{\hat P^I_{0 0}}{0}.\label{eq:ph4}
\end{eqnarray}
From $(\ref{eq:ph3})=(\ref{eq:ph4})$,
\begin{eqnarray}
	\frac {\bracket{0}{X_{n_{+1}} \hat P^I_{11}X_{n_{+1}}^\dagger }{0} }
	{ \bracket{0}{\hat P^I_{0 0}}{0} } 
		\simeq \frac {V_v(+ \; n_{-1} )}{2 V_v(n_{+1}\; -)}= \frac{1}{2}.
\end{eqnarray}
Here we used the relation $V_v(n_{-1}\;+) = V_v(+ \; n_{-1})=V_v(n_{+1}\; -)= V_v(- \; n_{+1}) $ . As the result, $\Delta E$ can be given analytically
\begin{eqnarray}
	\Delta E   &\simeq& -I(I+1) \sum_{n_{K=1}} \frac{4 V_v^2(n_{+1}\;- )}{ \omega_n a_0^2} \; \frac{ \bracket{0}{\hat P^I_{0 0}}{0} }{ \bracket{0}{X_{n_{+1}} \hat P^I_{11} X^\dagger_{n_{+1}}}{0} }\\
	&\simeq& -8I(I+1) a_0^{-2} \sum_{n_{K=1}} 
\omega_n^{-1} V_v^2(n_{+1}\;-).
\end{eqnarray}

\section{Evaluation of the $B(E2)$ and $\bracket{0}{\hat P^I}{0}$ }
 $B(E2)$ transition probability is given by 
\begin{eqnarray}
	B(E2;I\!+\!2 \rightarrow I)&\equiv& \frac{1}{2I+5}
		|\rbracket{I\!+\!2}{\hat Q_2}{I}|^2 .
\end{eqnarray}
where $\rbracket{I\!+\!2}{\hat Q_2}{I}$ is {\it the reduced matrix element} derived from {\it the Wigner-Eckart theorem},
\begin{eqnarray}
	\bracket{I\!+\! 2 \: M\!+\!\lambda}{\hat Q_{2\lambda}}{IM}= (2I+1)^{-1/2} \: \CG{I\!+\!2\; M\!+\!\lambda \: ; 2\; -\lambda}{I\: M} \rbracket{I\!+\!2}{\hat Q_2}{I}.	\nonumber\\
\end{eqnarray}
We have only to calculate $\bracket{I\!+\! 2 \: M}{\hat Q_{20}}{IM}$. In the space $\ket{IM}={\mathcal N}_0 \hat P^I_{M0}\ket{0}$
\begin{eqnarray}
	&&\bracket{I\!+\!2\:M}{\hat Q_{20}}{I M}\nonumber\\
%
	&=& \frac{ {\mathcal N}^{I+2}_0}{{\mathcal N}^I_0}(-1)^M \CG{I M;20}{I\!+\!2 \; -M}\sum_{\lambda=-2}^2  \CG{I\!-\!\lambda ;2\;\lambda}{I\!+\!2 \;0} g_\lambda (I),\quad 
\label{eq:q20}
\end{eqnarray}
where
\begin{eqnarray}
	g_\lambda (I)\equiv (-1)^\lambda \frac{ \bracket{0}{ \hat P^I_{0\lambda}\hat Q_{2\lambda} }{0} }{ \bracket{0}{ \hat P^I_{00} }{0} }.
\end{eqnarray}	
Here we used the relation
\begin{eqnarray}
	\hat P^I_{KM} \hat Q_{2\lambda}= \sum_{\lambda' I'M'K'}\ \CG{I'M';2\lambda}{IM}\CG{I'K';2\lambda'}{IK} \hat Q_{2\lambda'}\hat P^{I'}_{K'M'}.
\end{eqnarray}
Finally we can rewrite the $B(E2)$
\begin{eqnarray}
	B(E2;I\!+\! 2 \rightarrow I)= 
			\left[ \frac{ {\mathcal N}^{I+2}_0 }{{\mathcal N}^I_0}
\sum_\lambda \CG{I\;-\!\lambda;2\;\lambda}{I\!+\!2\;0} \; g_\lambda (I) \right]^2,
\end{eqnarray}
$g_\lambda (I)$ and ${\mathcal N}^{I+2}_0/{\mathcal N}^I_0$ can be derived approximately. Pair operator representation of $\hat Q_{2 \lambda} \ket{0}$ is
\begin{equation}
	\hat Q_{2 \lambda} \ket{0} =  \left\{
\begin{array}{ll}
	 \left( \left \langle \hat Q_{20} \right \rangle +\sum_n \xi_n X_n^\dagger \right) \ket{0} & (\mbox{for} \quad\lambda=0)\\
	 \left( \xi_\pm X^\dagger_{\pm} +  \sum_n \xi_n X_n^\dagger \right) \ket{0} & (\mbox{for} \quad\lambda=\pm1) \\
	  \sum_n \xi_n X_n^\dagger \ket{0} &  (\mbox{for} \quad\lambda=\pm2)
\end{array}\right..\qquad
\end{equation}
From our approximation, $g_\lambda (I)$ can be analytically written.
\begin{equation}
	g_\lambda (I) \simeq  \left\{
\begin{array}{ll}
	\left \langle \hat Q_{20} \right \rangle & (\mbox{for} \quad\lambda=0)\\
	-\xi_\pm a_0^{-1} \sqrt{I(I+1)} & (\mbox{for} \quad\lambda=\pm 1)\quad .\\
	 0 & (\mbox{for} \quad\lambda=\pm 2)
\end{array}\right.
\end{equation}
On the other hand, ${\mathcal N}^{I+2}_0 /{\mathcal N}^I_0$ is derived from $\bracket{0}{\hat P^{I\!+\!2}_{0M} \: \hat Q_{20}\:  \hat P^I_{M0}}{0}$
\begin{eqnarray}
	&&\bracket{0}{\hat P^{I\!+\!2}_{0M} \: \hat Q_{20}\:  \hat P^I_{M0}}{0}\nonumber\\
	&=& \CG{IM;20}{I\!+\!2 \:M} \bracket{0}{P^I_{00}}{0} \sum_{\lambda}
	\CG{I\: \lambda; 2 \;-\!\lambda}{I\!+\!2 \: 0} g_{\lambda}(I)\label{eq:pq1}\\
	&=& \CG{I\!+\!2 \:M ;20}{IM} \bracket{0}{P^{I+2}_{00}}{0} \sum_{\lambda}
	\CG{I\!+\!2 \: \lambda \: ; 2 \; -\!\lambda}{I0} g_{\lambda}(I+2)\label{eq:pq2}.
\end{eqnarray} 
From $(\ref{eq:pq1})=(\ref{eq:pq2})$, $(\ref{eq:Nratio})$ can be analytically given.

\end{document}